\documentclass[12pt,a4paper]{article}
\pdfoutput=1
\usepackage{jheppub}
\usepackage{caption} 
\usepackage{subcaption}
\usepackage[utf8]{inputenc}
\usepackage[T1]{fontenc}
\usepackage{epsfig}
\usepackage{multirow}
\usepackage{amssymb,amsfonts,amsmath}
\usepackage{graphicx}
\usepackage{breqn}
 
\font\cmss=cmss10
\font\cmsss=cmss10 at 7pt
\font\manual=manfnt

\newcommand{\bi}{\begin{itemize}}
\newcommand{\ei}{\end{itemize}}

\newcommand{\bea}{\begin{eqnarray}}
\newcommand{\eea}{\end{eqnarray}}
\newcommand{\be}{\begin{equation}}
\newcommand{\ee}{\end{equation}}
\newcommand{\ben}{\begin{eqnarray*}}
\newcommand{\een}{\end{eqnarray*}}
\newcommand{\bem}{\begin{pmatrix}}
\newcommand{\eem}{\end{pmatrix}}
\newcommand{\bl}{\begin{align}}
\newcommand{\el}{\end{align}}
\newcommand{\beg}{\begin{gather}}
\newcommand{\eeg}{\end{gather}}

\newcommand{\ve}{\varepsilon}

\newcommand{\tG}{{\tilde G}}

\newcommand{\tGa}{{\tilde \Gamma}}

\newcommand{\tS}{{\tilde \Sigma}}
\newcommand{\te}{{\tilde \eta}}

\newcommand{\IH}{\mathbb{H}}

\renewcommand{\b}{\beta}
\renewcommand{\d}{\delta}

\newcommand{\g}{\gamma}
\newcommand{\h}{\eta}

\renewcommand{\k}{\kappa}             

\newcommand{\m}{\mu}
\newcommand{\n}{\nu}

\renewcommand{\r}{\rho}                                     
\newcommand{\s}{\sigma}                                   
\renewcommand{\t}{\tau}

\newcommand{\G}{\Gamma}

\renewcommand{\L}{\Lambda}
\renewcommand{\O}{\Omega}

\newcommand{\half}{\frac{1}{2}}

\newcommand{\nn}{\nonumber}

\def\dbend{\lower3.5pt\hbox{\manual\char127}}

\def\IL{\relax{\rm I\kern-.18em L}}
\def\IH{\relax{\rm I\kern-.18em H}}
\def\rlx{\relax\leavevmode}

\def\ZZ{\rlx\leavevmode\ifmmode\mathchoice{\hbox{\cmss Z\kern-.4em Z}}
 {\hbox{\cmss Z\kern-.4em Z}}{\lower.9pt\hbox{\cmsss Z\kern-.36em Z}}
 {\lower1.2pt\hbox{\cmsss Z\kern-.36em Z}}\else{\cmss Z\kern-.4em
 Z}\fi}

\title{Quantum Cosmology in Four Dimensions}

\author[1, 2]{Teresa Bautista,}
\author[3, 4]{Andr\'e Benevides,}
\author[3, 1, 2]{Atish Dabholkar,}
\author[5]{Akash Goel}

\affiliation[1]{Sorbonne Universit\'es, UPMC Univ Paris 06\\
  UMR 7589, LPTHE, Paris, F-75005 France}
  
\affiliation[2]{CNRS, UMR 7589, LPTHE,  Paris, F-75005 France}  
  
\affiliation[3]{International Centre for Theoretical Physics\\
ICTP-UNESCO, Strada Costiera 11, Trieste 34151 Italy}

\affiliation[4]{SISSA, Via Bonomea 265, Trieste 34136 Italy}

\affiliation[5]{Indian Institute of Technology Kanpur, Kanpur 208016 India}

\abstract{We analyze the cosmological solutions to  the recently proposed nonlocal quantum effective  action for gravity with a cosmological term. We show that  the vacuum energy decays  with a  slow-roll parameter proportional to the anomalous gravitational dressings.}


\begin{document}
\maketitle

\section{Introduction}

It was pointed out  recently \cite{Dabholkar:2015qhk, Dabholkar:2015b} that Weyl anomalies in the renormalized quantum effective action of gravity can have important consequences for the gravitational dynamics on cosmological scales.  The relevant Weyl anomalies can be computed explicitly in a two-dimensional model of gravity and are summarized by a  nonlocal quantum effective action \cite{Bautista:2015wqy}. The resulting quantum  momentum tensor is nonlocal in general but simplifies for an isotropic and homogeneous universe and the cosmological equations can be solved analytically. The scale factor exhibits power law expansion driven entirely by the slowly decaying vacuum energy \cite{Dabholkar:2015qhk, Bautista:2015wqy}.

A four-dimensional phenomenological action \eqref{action} was proposed in \cite{Dabholkar:2015qhk, Dabholkar:2015b} motivated by these two-dimensional results  and from considerations of the local renormalization group \cite{Osborn:1989td,Jack:2013sha,Osborn:1991gm,Baume:2014rla}. This nonlocal  action parametrizes possible Weyl anomalies arising from the renormalization of composite operators. The resulting integro-differential equations describe the effective \textit{classical} dynamics of the spacetime metric  at long distances.

The nonlocality may be surprising at first but one must bear in mind  that the action \eqref{action} should be regarded as the 1PI effective action and not the Wilsonian effective action. In general, the 1PI effective action is expected to be nonlocal.  A class of non-local generalizations of the Einstein-Hilbert term have been considered earlier to analyze cosmological evolution \cite{Deser:2007jk,Deffayet:2009ca,Deser:2013uya, Tsamis:2014hra,Woodard:2014iga,Donoghue:2014yha,Woodard:2014wia} and can lead to interesting cosmology \cite{Park:2012cp,Barreira:2014kra}. The nonlocality that we consider is of a very specific kind and is constrained by the requirement that the action  should be a solution of the local renormalization group equation. The main idea is that anomalous dimensions of composite operators  modify the trace of the quantum momentum tensor as encapsulated by the local renormalization group equation. Therefore the  momentum tensor itself must be modified, which in turn can modify the gravitational dynamics.  This reasoning suggests a specific nonlocal generalization of Einstein Gravity.

In this paper we analyze the cosmological consequences of this action in a Robertson-Walker spacetime.  Somewhat surprisingly, the effective classical dynamics  can be solved analytically even in four dimensions   using the Weyl-invariant formulation discussed in \cite{Dabholkar:2015qhk,Dabholkar:2015b}. Our main conclusion is that nonzero anomalous gravitational dressings lead to a slow quantum decay of vacuum energy just as in two dimensions.  

The paper is organized as follows. In \S\ref{Dynamics} we review the nonlocal four-dimensional action proposed in \cite{Dabholkar:2015qhk,Dabholkar:2015b} and  derive the cosmological evolution equations for an isotropic and homogeneous universe.  
In \S\ref{Cosmology} we present the solutions to these equations which describe  an expanding  universe with a  decaying vacuum energy density.
We give estimates of the anomalous gravitational dressings and conclude with a  discussion of  the theoretical and cosmological implications. 

\section{Quantum  Gravity at Long Distances\label{Dynamics}}

We are interested in the quantum effective action for the metric obtained by integrating out the quantum fluctuations of various fields valid at distances large compared to the Planck distance. The essential lesson that emerges from the study of the two-dimensional model \cite{Dabholkar:2015qhk,Bautista:2015wqy} is that the physical coupling constants are the couplings of the {gravitationally dressed} operators. The anomalous dimensions of the   dressed operators are  in principle different from the anomalous dimensions of the undressed operators. This applies in particular to the square-root of the determinant of the metric corresponding to the cosmological term as well as to the Einstein-Hilbert term. The quantum effective action \eqref{action} should take into account these anomalous gravitational dressings and can be obtained either by using the background field method or as a solution of the local renormalization group equation \cite{Dabholkar:2015b}. 

\subsection{A Nonlocal Action for Gravity \label{Physical}}

The nonlocal  action   in the physical gauge is given by:
\begin{equation}\label{action}
I_{G}[g]=\frac{M_p^2}{2}\,\int d^4 x\sqrt{-g}\, \left( R[g] \, e^{-\Gamma_{K}(\Sigma_g)} 
- 2\, \Lambda\,  e^{-\Gamma_{\Lambda}(\Sigma_g)} \right)
\end{equation}
where $M_p$ is the reduced Planck mass and the $\Gamma_i(\Sigma_g)$, $i=K,\L$ are  the integrated anomalous gravitational dressings. 
The field $\Sigma_{g}$   is a nonlocal functional of the metric $g_{\m\n}$ defined by  \cite{Fradkin:1978yf, Paneitz:2008, Riegert:1984kt}
\begin{equation}\label{Sigma}
 \Sigma_g(x) = \frac{1}{4}\int\! d^4 y\,\sqrt{-g}\, G_4(x,y)\, F_4[g](y) \, ,
\end{equation}
where
\begin{align}\label{F4}
F_4[g] = E_4[g]-\frac{2}{3} \nabla^2 R[g]\, , \qquad
 E_4[g] =(R^{\m\n\r\s} R_{\m\n\r\s} - 4 R^{\m\n} R_{\m\n} + R^2)[g] \, ;
\end{align}
and $G_{4}(x,y)$ is the Green function of the Weyl covariant quartic differential operator 
\begin{equation}\label{diff-op}
 \Delta_4[g]= \left( \nabla^2 \right)^2 + 2 R^{\m\n}  \nabla_{\m}\nabla_{\n} + \frac{1}{3}\left(\nabla^{\n}R\right)\nabla_{\n} - \frac{2}{3}R\, \nabla^2 \, 
\end{equation}
on the $g_{\m\n}$ background satisfying
\be\label{green-equation}
 \Delta^{x}_4[g] G_{4}(x, y) = \d^{(4)}(x, y) :=\frac{ \d^{(4)}(x -y) }{\sqrt{-g}}\, .
\ee
For metrics  related by a Weyl rescaling 
\be\label{g-reference}
g_{\m\n}=e^{2\Sigma_g(x)}\, {\bar \eta}_{\m\n}\, ,
\ee
the scalars $F_{4}$ are related by
\begin{equation}\label{ftrans}
 F_4[ g] = e^{-4\Sigma_g(x)} \left( F_4[\bar \eta] +4  \, \Delta_4 [\bar \eta]\, \Sigma_{g} \right)\, ,
\end{equation}
and the operators $\Delta_{4}$ are related by
\begin{equation}
\Delta_4[g]= e^{-4\Sigma_{g}}\, \Delta_4[\bar \eta] \, .
\end{equation}
One can choose a gauge in which $\Sigma_g(x)$ becomes the conformal factor of the metric with respect to a reference metric  ${\bar \eta}_{\m\n}$ which satisfies the F-flatness condition
$F_4[\bar \eta]=0$. The expression \eqref{Sigma} is obtained in this F-flat gauge by  inverting \eqref{ftrans}. Note that the action \eqref{action} should be regarded as the in-in effective action and hence one must impose  retarded boundary conditions. This ensures that the propagation is causal.  

We  emphasize that the action \eqref{action} is the result of having performed a path integral and is not to be quantized further but is to  be used for studying the effective classical dynamics.
In principle,  the dressing functions can be computed  \textit{ab initio} in a given microscopic theory  \cite{Dabholkar:2015b}. For now, we view these functions as a phenomenological parametrization of possible Weyl anomalies and analyze the cosmological solutions of \eqref{action} in terms of these functions. 

\subsection{Weyl-Invariant Nonlocal Action \label{Weyl}}

The variation  of \eqref{action} with respect to $g_{\m\n}$ is very cumbersome because both $\Delta_{4}$ and $F_{4}$ have a complicated dependence on the metric.  The Weyl-invariant formulation discussed in \cite{Dabholkar:2015qhk,Dabholkar:2015b}  leads to considerable simplification  by exploiting the fact that the spatially flat Robertson-Walker metric  is Weyl equivalent to the flat Minkowski metric. 

In the Weyl-invariant formulation, one introduces a Weyl compensator field $\O(x)$ and a fiducial metric $h_{\m\n}$ to write the physical metric as
\be\label{Weyl-split}
g_{\m\n}= e^{2\Omega(x)} h_{\m\n}\, .
\ee
With this arbitrary split, the physical metric is invariant under a `fiducial Weyl transformation' 
\be\label{Weyl-transf}
h_{\m\n}\rightarrow e^{2\xi(x)} h_{\m\n}\, ,\qquad \O(x)\rightarrow \O(x)-\xi(x)\,  .
\ee
The fiducial metric  can be further parametrized in terms of  an F-flat reference metric $\bar \h_{\m\n}$ as
\be
h_{\m\n}= e^{2\Sigma_h(x)} \bar \eta_{\m\n} \, .
\ee
Then,  the Weyl factor $\Sigma_h(x)$ is given by \eqref{Sigma} evaluated on $h_{\m\n}$. Equation \eqref{Weyl-split} implies
\be\label{Sigma-Omega}
\Sigma_g=\O+\Sigma_h\, 
\ee
and \eqref{Weyl-transf} implies  that $\Sigma_g(x)$  is invariant under the fiducial Weyl symmetry.  Let $M_{0}$ be the UV cutoff scale below the Planck scale. One can then define the gravitational coupling $\kappa^{2}$ by
\be
M_{p}^{2} = \frac{M_{0}^{2}}{\kappa^{2}} \, .
\ee
Henceforth we choose units so that $M_{0} =1$. 
Substituting \eqref{Weyl-split} and \eqref{Sigma-Omega} into the action \eqref{action}, one obtains the Weyl-invariant quantum action 
\be
I_G[h,\O]=\frac{1}{2\k^{2}}\,\int d^4x \sqrt{-h} \, e^{4\Omega} \left(R[h e^{2\Omega}] \, e^{-\Gamma_K(\Omega+\Sigma_h )}   - 2\, \Lambda\, e^{-\Gamma_{\Lambda}(\Omega+\Sigma_h )} \right) \, .
\ee
Using the Weyl transformation of the Ricci scalar and  integrating by parts, we obtain
\begin{dmath}\label{Waction}
I_G[h,\O]=\frac{1}{2\k^{2}}\int dx  \,\left[ \left(R[h]+ 6\,(1-\Gamma_K^{(1)})\,|\nabla\Omega|^2 -\, 6 \,\Gamma_K^{(1)} \,\nabla\Omega \cdot \nabla\Sigma_h
\right)e^{2\Omega-\Gamma_K} - 2\, \Lambda\, e^{4\Omega-\Gamma_{\Lambda}}  \right]\, 
\end{dmath}
where  $dx\equiv d^4x\,\sqrt{-h}$ and\footnote{Henceforth, all covariant derivatives and contractions are with respect to the fiducial metric $h_{\m\n}$.} $\G_i^{(n)}(\O+\Sigma_h)$ are the $n$-th derivatives of the dressing functions. 

The action \eqref{Waction}  now has an enlarged gauge symmetry  that includes  local Weyl symmetry in addition to diffeomorphisms.  We have introduced an additional scalar degree of freedom in the process, but  the number of physical degrees of freedom remains unchanged. Using the fiducial Weyl invariance one can  choose a `physical gauge'  $\Omega =0$ so that the physical metric is identified with the fiducial metric and one recovers the action \eqref{action}. Alternatively, one can keep $\Omega$ arbitrary and impose a scalar gauge condition on the fiducial metric such as  $F_{4}[h] =0$. In this F-flat gauge 
 $\Sigma_{h}=0$ and $\Sigma_{g} = \O$. 
 
\subsection{Evolution Equations for Cosmology \label{F-flat}}

A homogeneous and isotropic universe is described by the Robertson-Walker metric. For simplicity we consider the case when the spatial section is flat. In this case, one  can choose a gauge    in which the fiducial metric $h_{\m\n}$ equals the flat Minkowski metric $\eta_{\m\n}$. Since this is determined purely by the symmetry of the problem, the entire dynamics now resides in the Weyl compensator. Moreover,   the Minkowski metric is not only F-flat but Riemann-flat. Consequently the variation of the nonlocal terms in the action \eqref{Waction} simplifies considerably.

Since $F_{4}[\eta] = 0$, the variation of the Green function does not contribute, and we obtain 
\be
\d\Sigma_h(x)=\frac{1}{4}\int dy \, G_4(x,y)\, \d F_4[h](y)\, .
\ee
Furthermore,  the quadratic terms involving the Riemann curvature tensors do not contribute to the variation of $F_{4}[h]$ when evaluated around $\eta_{\m\n}$ . The only nonzero variation comes from the variations of the term linear in the curvatures:
\be
\d F_4[h](y)= -\frac{2}{3}\, \nabla^2\, \d R[h]\, .
\ee
The total variation of $\Sigma_h$ after an integration by parts is then given by
\be\label{Sigma-variation}
\d\Sigma_h(x)=\frac{1}{6}\int dy \, \d h^{\m\n}(y)\, (\nabla_\m\nabla_\n-h_{\m\n} \nabla^2) \, \nabla^2\,G_4(x,y)\,.
\ee
After  performing the variation in the fiducial frame, it is convenient to rewrite the equations of motion in terms of the gauge-invariant physical metric using \eqref{Weyl-split}. Under a Weyl transformation, the Einstein tensor transforms as \cite{Dabholkar:2015b}
\bea\label{identities}
E_{\m\n}[g]&=&E_{\m\n}[h] + D_{\m\n}[h,\O]\, , \nn\\
D_{\m\n}[h,\O]&:=&-2\left(\nabla_\m\nabla_\n-h_{\m\n}\nabla^2\right)\O+2\left(\nabla_\m\O\nabla_\n\O+\half h_{\m\n} |\nabla\O|^2\right) \, .
\eea
Substituting the above in the variation of \eqref{Waction} yields the  equations of motion for the physical metric 
\be\label{QEinstein}
E_{\m\n}[g]=\k^2 \,(T^K_{\m\n}+T^\L_{\m\n})
\ee
where $T^K_{\m\n}$ is the  momentum tensor of  the `gravifluid' \cite{Dabholkar:2015b}, of purely geometric origin
\bea\label{gravi-tensor}
\k^2\, T^K_{\m\n}(x) &=&2\, \G_K^{(1)} \left( \nabla_\m \O\,\nabla_\n\O+\half \h_{\m\n} |\nabla\O|^2\right)\\
&&\,+ \left( ({\G_K^{(1)}})^2-\G_K^{(2)}\right)  \left( \nabla_\m \O\,\nabla_\n\O- \h_{\m\n} |\nabla\O|^2\right)
-\G_K^{(1)} \left(\nabla_\m\nabla_\n -\h_{\m\n} \nabla^2\right) \O \nn\\
&&\,- \,e^{-2\O+\G_K}\,\int dy \, \G_K^{(1)} e^{2\O-\G_K}\,(\nabla^2\O+|\nabla\O|^2) \,\left(\nabla_\m\nabla_\n-\h_{\m\n} \nabla^2\right) \nabla^2 G_4(x,y)\, ;\nn
\eea
and $T^\L_{\m\n}$ is the momentum tensor of the `vacuum fluid' 
\bea\label{cosmo-tensor}
\k^2 \,T^\L_{\m\n}(x) &=& - \,\L \,\h_{\m\n} \,e^{2\O+\G_K-\G_\L}\\
&&\,- \,\frac{\L}{3}\, e^{-2\O+\G_K}\,\int dy \, \G_\L^{(1)}\, e^{4\O-\G_\L}\,\left(\nabla_\m\nabla_\n-\h_{\m\n} \nabla^2\right) \nabla^2 G_4(x,y)\, .\nn
\eea
We emphasize that  the contribution from the `gravifluid' is purely geometric in origin and in principle  belongs to the left hand side of the equation \eqref{QEinstein} on the same footing as the Einstein tensor.
Since the  fiducial metric is flat in this context, the equation \eqref{QEinstein} reduces to 
\be\label{Dequation}
D_{\m\n}[\h,\O] = \k^2 \left(T^K_{\m\n}+\,T^\L_{\m\n} \right)\, .
\ee

\subsection{Cosmological  Equations in an Alternative Gauge \label{R-flat}}

As described in \cite{Dabholkar:2015b}, it is possible to choose an alternative gauge in which the conformal factor $\tS_g(x)$ is defined with respect to an R-flat reference metric $\te_{\m\n}$ \cite{Fradkin:1978yw, Barvinsky:1995it}. In the  R-flat gauge,  the expression for the conformal factor follows from the Weyl transformation of the Ricci scalar
\be
R[g]=e^{-2\tS_g}\,\left( R[\te]-6 \nabla^2 \tS_g -6 |\nabla \tS_g|^2\right)\, .
\ee
Imposing $R[\te]=0$ and inverting the above equation gives 
\be\label{Sigma-tilde}
\tS_g(x)=-\,\ln\left( 1-\int d^dy\,\sqrt{-g}\, \,\tG(x,y)\, R[g](y)\right)\, ,
\ee
where $\tG(x,y)$ is the Green function of the differential operator 
\be\label{Green-tilde}
\left(-6\,\nabla^2+R\right)^x[g] \,\tG(x,y)=\d^{(4)}(x,y)\, .
\ee
To use this gauge in the quantum action, we express the fiducial metric as
\be
h_{\m\n}=e^{2\tS_h}\, \te_{\m\n}\, ,\qquad \tS_g=\O+\tS_h\, .
\ee
We introduce the Weyl split \eqref{Weyl-split}  into the analog of the action \eqref{action}  in the R-flat gauge with gravitational dressing functions\footnote{Note that $ \tilde\eta_{\m\n} =e^{2\Sigma_{\tilde\eta} }\, \bar\eta_{\m\n}$ and  $ \Sigma_{g} = \tilde \Sigma_{g} + \Sigma_{\tilde\eta} $. As a result,  the integrated anomalous gravitational dressing functions in the two gauges are related by a shift: $\tilde \Gamma_{i}(\tilde \Sigma_{g})  = \Gamma_{i}(\tilde \Sigma_{g}+ \Sigma_{\tilde\eta} )$.} $\tGa_i(\tS_g)$. The  Weyl-invariant action then becomes\begin{dmath}\label{RWaction}
I_G[h,\O]=\frac{1}{2\k^{2}}\int dx  \,\left[ \left(R[h]+ 6\,(1-\tGa_K^{(1)})\,|\nabla\Omega|^2 -\, 6 \,\tGa_K^{(1)} \,\nabla\Omega \cdot \nabla\tS_h
\right)e^{2\Omega-\tGa_K} - 2\, \Lambda\, e^{4\Omega-\tGa_{\Lambda}}  \right]\,
\end{dmath}
where now $\tGa_i=\tGa_i(\O+\tS_h)$.

The equations of motion for a spatially-flat Robertson-Walker spacetime then follow from the variation of this action around $h_{\m\n}=\eta_{\m\n}$. On a flat background, the variation of $\tS_h$ \eqref{Sigma-tilde}  receives no contribution from the variation of the Green function and is given by
\be\label{tSigma-variation}
\d \tS_h=  \int d^4y\, \tG(x,y)\, \d\, (\sqrt{-h}\, R_h(y) )= \int dy\, \d h^{\m\n} \,\left( -\nabla_\m\nabla_\n + h_{\m\n} \nabla^2 \right)\, \tG(x,y)\, .
\ee
Furthermore,  since $R_{\eta}=0$, the equation \eqref{Green-tilde} becomes the  Green equation for the flat Laplacian.
Comparing with the Green equation \eqref{green-equation} of $\Delta_4$ on a flat background, we find that the two Green functions are related through
\be
\tG(x,y)=-\frac{1}{6} \,\nabla^2 G_4(x,y)\, .
\ee
Introducing this in \eqref{tSigma-variation} we recover the same variation of $\Sigma_h(x)$ in a flat background \eqref{Sigma-variation}:
\be
\d \tS_h=\frac{1}{6}\int dy \, \d h^{\m\n}\, (\nabla_\m\nabla_\n-h_{\m\n} \nabla^2) \, \nabla^2\,G_4(x,y)\,.
\ee
Since $h_{\m\n}$ is taken to be Minkowski, $\Sigma_h=\tS_h=0$, and therefore  $\G_i(\O)=\tGa_i(\O)$. As a result, the equations of motion obtained in the two gauges are identical. 

\section{Quantum Decay of Vacuum Energy \label{Cosmology}}

Thus far we have only required that the physical metric be Weyl equivalent to the Minkowski metric. The  equations derived in the previous section are valid  generally as long as the Weyl tensor of the physical metric vanishes.  In a spatially-flat Robertson-Walker spacetime  there is further simplification because the  scale factor of the physical metric 
is a function of  only the conformal time\footnote{Conformal time $\t$ is related to comoving cosmological time $t$ by $d\tau = \frac{dt}{a(t)} $.} $\tau$. With our gauge choice of $h_{\m\n}=\eta_{\m\n}$, we can write \be
a(\tau) = e^{\O(\t)} \, .
\ee 
The momentum tensors \eqref{gravi-tensor} and \eqref{cosmo-tensor} now simplify further and the  integro-differential equations \eqref{Dequation} reduce to an ordinary differential equation of the usual Friedmann-Lemaître type but for an effective quantum fluid with an unusual equation of state.

\subsection{Effective Quantum Fluid}

For the Robertson-Walker metric, the explicit form of  $G_4(x,y)$ discussed in \S\ref{Green} is actually not needed because its contribution to the momentum tensors is of the form
\be
\int dy\, F[\O]\, \left(\nabla_\m\nabla_\n-h_{\m\n} \nabla^2\right) \nabla^2 G_4(x,y)\, .
\ee
The differential operator in the parenthesis vanishes when $\m=\n=\t$, so it does not contribute to the energy density. For all other components, the first term in the parenthesis vanishes after integration by parts, and for the components $\m,\n=i$, the second term in the parenthesis can be identified with the Green equation of $\Delta_4[\eta]$. 
It follows that the quantum momentum tensors \eqref{gravi-tensor} and \eqref{cosmo-tensor} correspond to perfect fluids, consistent with isotropy and homogeneity, although they are not separately conserved.

 The density and pressure of the vacuum fluid are given by
\bea
\rho_\L(t)&=& \frac{\L}{\k^2}\,e^{\G_K-\G_\L}\, , \, \quad p_\L(t) = w_{\L}(t) \, \rho_\L(t)\,, \label{v-density} \\
w_{\L}(t)&=&\left(-1+\frac{\G_\L^{(1)}}{3}\right)\,  .\label{v-index}
\eea 
The density and pressure of the gravifluid \textit{after  using} the equations of motion are given by 
\bea
\rho_K(t)&=&    \frac{\L}{\k^2}\,\frac{\G_K^{(1)}}{1-\G_K^{(1)}}\,e^{\G_K-\G_\L}\, , \quad p_K(t) = w_{K}(t) \, \rho_K(t) \, ,\label{K-density} \\
w_K(t)&=&\left( -1+\frac{\G_\L^{(1)}-1}{3}-\frac{\G_K^{(2)}}{3\,\G_K^{(1)}\,(1-\G_K^{(1)})}\right)\,  .\label{K-index}
\eea
We have written the expressions above in the `barotropic' form with the effective pressure proportional to the effective density,  but the anomalous gravitational dressings $\Gamma_i(\ln a(t))$ are in general nontrivial functions of the comoving time. As a result, the barotropic indices $w_{\L}$  and $w_{K}$ are in general \textit{time-dependent}\footnote{Recall that in classical cosmology the commonly encountered fluids have the barotropic index $-1$ for the cosmological constant,  $0$ for matter,  and $1/3$ for radiation.} and should be regarded as a convenient parametrization.

Combining the two contributions  one obtains the total momentum tensor on the right hand side of the equation \eqref{QEinstein}. It is a perfect fluid  with the effective density and pressure given by 
\bea
\rho_{e}(t)&=&    \frac{\L}{\k^2}\,\frac{1}{1-\G_K^{(1)}}\,e^{\G_K-\G_\L}\, , \quad p_e(t) = w_{e}(t) \, \rho_e(t) \label{e-density} \\
w_e(t)&=& \left(-1 + \frac{\g}{3}\right) \, , \quad \g = \left( \G_{\L}^{(1)} - \G_{K}^{(1)} -\frac{\G_K^{(2)}}{\,(1-\G_K^{(1)})} \right)\, . \label{e-index}
\eea
With this effective density, the equation of motion reduces to the first Friedmann equation
\bea\label{Friedmann1}
H^{2} =  \frac{ \kappa^{2} \r_{e} } {3}  \, 
\eea
where $H:= \dot{a}/a$ as usual. Note that our conclusions thus far follow purely from the symmetry considerations of isotropy, homogeneity, and spatial flatness. 

The momentum tensor for the gravifluid \eqref{K-density} is proportional to the cosmological constant after using the equations of motion in  a spatially-flat Robertson-Walker spacetime.  As a result the total momentum tensor for the effective fluid is proportional to the cosmological constant. This implies that in the absence of the cosmological constant,   the Minkowski metric  continues to be an exact solution of the  new equations \eqref{QEinstein} in vacuum. On the other hand, for positive cosmological constant,  the classical de Sitter solution is no longer a solution of the quantum equations \eqref{QEinstein} as we describe below.

The conservation equation for the effective fluid is 
\be\label{Conservation}
\dot{\rho_{e}}= - 3 (p_{e} + \r_{e}) H \, .
\ee
 A useful consistency check is that the expressions \eqref{e-density} and \eqref{e-index} satisfy the conservation equation. It is of course guaranteed by the fact that the nonlocal action \eqref{action} is coordinate invariant and follows from the Bianchi identity.  Note, however,  that the gravifluid and the vacuum fluid are not conserved separately  for nonzero $\G_K$.

\subsection{Cosmology of the Decaying Vacuum Energy}

The expressions \eqref{e-density} and \eqref{e-index} for the effective density and pressure already give their functional dependence on the scale factor. As discussed above, they automatically solve the conservation equation \eqref{Conservation}. 
Our task is then reduced to solving the equation \eqref{Friedmann1} to obtain the scale factor as a function of the cosmological time. Even though \eqref{Friedmann1} is much simpler than an integro-differential equation, it is nevertheless a complicated ordinary differential equation. In general,  the  integrated anomalous dressings $\G_{K}$ and $\G_{\L}$ are nontrivial functions of the scale factor and this equation can be solved only numerically.

Analytic solutions are possible when  $\G_{K}$ and $\G_{\L}$ are both linear functions of $\Omega$:
\be
\G_K(\O)=\g_K\, \O(x), \qquad \G_\L(\O)=\g_\L\,\O(x)\, ,
\ee
where $\g_i$ are constants\footnote{ In two dimensions,  $\G_K(\O)$ and $\G_\L(\O)$   are indeed linear functions with  $\g_{K}=0$ and $\g_{\L}=2\b^{2}$ \cite{Bautista:2015wqy}. This is a consequence of the conformal invariance of the timelike Liouville theory. In general, this need not be true, but it may be possible to approximate  the integrated anomalous gravitational dressings  by linear functions for long enough time intervals during the evolution of the universe.}.  The barotropic index for both the vacuum fluid \eqref{v-index} and the gravifluid \eqref{K-index} becomes constant. It is useful to consider this case to gain some understanding of the resulting solutions. From \eqref{v-index} and \eqref{K-index} we obtain
\be
w_\L=-1+\frac{\g_\L}{3}\qquad w_K=-1+\frac{\g_\L-1}{3}\, .
\ee
More interestingly, the effective fluid appearing on the right-hand side of the Einstein equations becomes also barotropic with index
\be\label{barox}
w_{e}=-1+\frac{\g}{3}\, ,  \quad \textrm{with} \quad \g =\g_\L-\g_K \, .
\ee
The cosmological solution to \eqref{QEinstein}  is then given by
\be\label{modified-deSitter}
\rho_e(t)  = \rho_{e*} \left(\frac{a}{a_*}\right)^{-\g} \, , \qquad  a(t) = a_{*}(1 + \frac{\gamma}{2} H_{*} t) ^{\frac{2}{\g}} \, ,
\ee
where $\r_{e*}$, $H_{*}$, $a_{*}$ are the initial values at the beginning of universe at time $t=0$. The densities of the vacuum fluid and the gravifluid are given by
\be\label{evol-densities}
\r_\L(t)=(1-\g_K)\, \r_e(t)  \, ,\qquad \r_K(t)=\g_K\, \r_e(t)\, .
\ee
The conformal time as  a function of the comoving time is given by
\be\label{conformal-time}
\t = \t_{*} (1 + \frac{\gamma}{2} H_{*} t)^{\frac{\g -2}{\g}} \, \quad \textrm{where} \quad \t_{*} :=  \frac{2}{(\g-2) H_{*} a_{*}} \, . 
\ee
The range of  $t$ is  $0 \leq t < \infty$ with the universe starting with scale factor $a_{*}$. The range of $\tau$ is 
\bea
|\t_{*}| \leq \t <  \infty &\quad& \textrm{for} \quad \gamma >  2\, , \\
- |\t_{*}| \leq \t <  0 &\quad& \textrm{for} \quad 0 \leq \gamma < 2 \, .
\eea
As a function of $\t$, the scale factor and the density are given by 
\be\label{conformal-solution}
a(\t) = a_{*} (\frac{\t}{\t_{*}})^{\frac{2}{\g -2}} \, , \quad \rho_{e}(\t) = \rho_{e*} (\frac{\t}{\t_{*}})^{-\frac{2\g}{\g -2}} \, .
\ee

In the semiclassical approximation both anomalous dressings are expected to be small. A spatially-flat Robertson-Walker solution is only compatible with $\L \geq 0$.   For positive $\g$,  our model describes an expanding universe driven by an effective fluid with a barotropic index that is slightly larger than its classical value $-1$.   We arrive at  the conclusion that the vacuum energy density decays from its initial value $\r_{e*}$ which could be of the order of the string scale or the scale of supersymmetry breaking. The classical exponential expansion of de Sitter spacetime is  slowed down to a power law expansion as a result of the quantum anomalous gravitational dressings. In the limit of vanishing $\g$, one recovers de Sitter spacetime with constant density. 

The quantum decay of vacuum energy and the dynamics of the Omega field provides a new mechanism to   drive slow-roll inflation in the early universe.
It is convenient to define  slow-roll parameters as usual in terms of fractional change in the Hubble parameter and its derivative:
\be
\ve_{H}:=-\frac{\dot{H}}{H^2}=-\frac{d\ln H}{Hdt} \quad \textrm{and} \quad \eta_{H}:=\frac{\dot \ve_{H}}{H\ve_{H}} =\frac{d\ln \ve_{H}}{Hdt} \, .
\ee
For the solution
\be
\ve_{H} = \frac{\gamma}{2} \quad \textrm{and} \quad \eta_{H} =0 \, .
\ee
The condition for accelerated expansion ($\ddot a>0$) requires that  $\ve_{H}$ should be less than one.  Slow-roll inflation requires that $\ve_{H} \ll 1$.  It is also necessary that $\h_{H} \ll 1$ so that inflation lasts long enough. 
Since $\g$ is small in the semiclassical approximation, all these conditions would be satisfied. A generic prediction is that  $\h_{H} =0 $.
Thus, the quantum decay of vaccum energy can drive slow-roll inflation in the early universe. For small $\g$, the scale factor expands almost exponentially as a power law with a very high exponent. Nonzero $\ve_{H}$ measures the deviation from exact exponential expansion but the parameter $\eta_{H}$ vanishes as in exact de Sitter spacetime. 

For a general functional form of the anomalous dressings $\Gamma_{K}$ and $\Gamma_{\L}$, the equation \eqref{Friedmann1} represents a novel generalization of the usual Friedmann equation because the equation of state of the effective fluid is rather unusual. It is conceivable that this has interesting consequences for  early cosmology. Numerical integration may be necessary to find the time-dependence of the scale factor. 
However, we see from \eqref{e-density} that as long as $\G_{\L} - \G_{K}$ is positive during the cosmological history, vacuum energy will decay. For negative  $\g$, the \textit{null energy condition} would be violated. In this case, the effective fluid could provide a model of phantom energy \cite{Weinberg:2008zzc}.

\subsection{Broken Time Translation Symmetry and Stability}

This novel mechanism for the decay of the vacuum energy raises the following puzzle. Unlike the classical de Sitter solution, our quantum corrected slow-roll solution \eqref{modified-deSitter} breaks the global time translation symmetry
\be
t\rightarrow t + \pi
\ee
 of the action \eqref{action} for a constant $\pi$.  If a solution breaks a global symmetry of an action, the symmetry-transform of a given solution generates a new solution. This implies that if one now considers a position dependent symmetry parameter $\pi(x)$ then the  effective action for $\pi(x)$ must be derivatively coupled so that there is a flat direction and  arbitrary constant $\pi$ is a solution of the equations of motion that follow from this effective action.  Correspondingly,  one expects a Nambu-Goldstone like  scalar fluctuation mode. In usual inflationary models this scalar mode can be identified with a gauge-invariant combination of the inflaton and the metric. This idea is the basis of effective field theories of inflation \cite{Cheung:2007st,Weinberg:2008hq}.  Where is this additional scalar degree of freedom? One could pose the puzzle slightly differently. Time translation symmetry is part of the diffeomorphism group. How can quantum effects break this symmetry?

The resolution of this puzzle is as follows. The scale factor of our solution has an initial value $a_{*}$ at the initial value surface $t=0$. Since we are using semiclassical gravity, $a_{*}$ can be taken to be of the order of  the short-distance cutoff scale a little larger than the Planck length. This means that, unlike the eternal de Sitter solution,  one cannot continue this solution to times earlier than $t=0$. The global time translation  symmetry is thus explicitly broken by the fact that one must cutoff the evolution with an initial value surface in the early universe and impose  initial conditions. Even though the action is invariant under the time translation symmetry, the initial conditions are not. Thus, one   cannot apply the argument above to generate new solutions from a given solution to deduce the existence of a propagating scalar degree of freedom. 

One can  state the result slightly differently. 
The nonlocal expression \eqref{Sigma}  for the Weyl factor follows from inverting  \eqref{ftrans} only if one discards all solutions
of the homogeneous equation
\be\label{homo}
\Delta_{4} \Sigma_{g} =0 \, .
\ee
These solutions correspond precisely to the would-be Nambu-Goldstone  scalar fluctuations.
The initial conditions  on $\Sigma_{g}$ on the initial value surface ensure that $\Sigma_{g}$ is determined entirely in terms of the metric and is not an additional propagating field.  

It is possible to reformulate the  argument above using a local action.
One can recast the nonlocal action \eqref{action} in a local form \cite{Nojiri:2007uq,Tsamis:2014hra} by introducing two auxiliary scalar fields $\Sigma(x)$ and $\Psi(x)$ with the action
\begin{equation}\label{local-action}
S[g,\Sigma,\psi]=\frac{M_p^2}{2}\,\int\! d^4 x\sqrt{-g}\, \left[  R[g] e^{-\Gamma_{K}(\Sigma)} 
- 2 \Lambda e^{-\Gamma_{\Lambda}(\Sigma)}+ \Psi\left( \Delta_4 \Sigma - \frac{1}{4}\,F_4 [g]\right) \right] \,. 
\end{equation}
The equations of motion for the two auxiliary fields are
\bea\label{emotion-aux}
\Delta_4 \Psi (x) =  R[g]\,\G_K^{(1)}\, e^{-\G_K(\Sigma)}-2\L\,\G_\L^{(1)}\, e^{-\G_\L(\Sigma)} \, ; \quad \quad
\Delta_4\Sigma (x) = \frac{1}{4}\,F_4 [g]\,.
\eea
The field $\Psi(x)$ acts therefore as a Lagrange multiplier for the condition $\Sigma=\Sigma_g$, and we recover \eqref{action} upon using its equation of motion in \eqref{local-action}.
This local action will reduce to the original nonlocal action only if  the homogeneous solutions of \eqref{emotion-aux} are eliminated by imposing an initial condition for $\Psi$ and $\Sigma$ that is similar to the initial condition for $\Sigma_{g}$ .  This ensures that the only propagating degrees of freedom are the usual tensor fluctuations  of the metric and there are no additional scalar fluctuations.

If a Lagrangian depends on higher time derivatives of the fields,  then one should also worry about the possibility of  the Ostrogradsky instability \cite{Woodard:2015zca}. 
We do not carry out the  stability analysis of our action in this paper but refer the reader to the stability analysis  for a  class of nonlocal actions \cite{Deser:2013uya,Tsamis:2014hra} similar to the one we consider in this paper in the R-flat gauge in \S\ref{R-flat}. 

These conclusions are physically reasonable  from the point of view of the original quantum path integral. The action \eqref{action} is the quantum 1PI-effective action for the background metric obtained by a semiclassical evaluation of  the path integral at weak coupling. It would be strange if one were to discover  an extra scalar degree of freedom or unphysical instability in this IR effective action if the starting point is a well-defined path integral.

\subsection{Quantum Gravity and De Sitter Spacetime \label{Discussion}}

The idea of  vacuum energy decay caused by infrared quantum effects has been explored earlier in four-dimensional gravity by several physicists. There is considerable divergence in the literature about the final result \cite{Polyakov:1982ug,Mottola:1984ar,Antoniadis:1985pj,Tsamis:1996qm,Tsamis:1996qq,Antoniadis:1991fa,Polyakov:2007mm,Polyakov:2012uc,Romania:2012av} and more generally about infrared effects in nearly de Sitter spacetime \cite{Starobinsky:1994bd,Weinberg:2005vy,Weinberg:2006ac,Senatore:2009cf,Kitamoto:2010si,Rajaraman:2010xd,
Kahya:2010xh,Marolf:2010zp,Giddings:2010nc,Higuchi:2011vw,Marolf:2011sh,Pimentel:2012tw, Akhmedov:2013vka}. 
One of the  new ingredients in the present work is to summarize the quantum effects in terms of a gauge-invariant  nonlocal action. This way of organizing the analysis may be useful for future explorations since it separates the computation of the anomalous dressings from the classical evolution. Since the effective action \eqref{action} is a solution to the local renormalization group equation, it   effectively sums up the leading logarithms.  To see this explicitly,  one can  expand the solution \eqref{conformal-solution} for the scale factor for small $\g$: 
\be\label{conformal-solution2}
a(\t) = a_{*} (\frac{\t}{\t_{*}})^{\frac{2}{\g -2}} =   a_{*} 
(\frac{\t_{*}}{\t}) \, (\frac{\t}{\t_{*}})^{\frac{-\g}{2-\g}} \sim \frac{a_{*} \tau_{*}}{\tau} \left[ 1 - \frac{\gamma}{2} \log{\left(\frac{\tau}{\tau_{*}}\right)} + \ldots \right] \, .
\ee
One obtains the usual de Sitter solution when  $\g=0$. For nonzero $\g$ there are logarithmic corrections which add up to  a small exponent which slows down the de Sitter expansion.

As it stands \eqref{action} should be regarded as a phenomenological parametrization  in terms of the integrated anomalous gravitational dressings. These anomalous dressings are computable in a given microscopic theory. Renormalization of Newton's constant and the cosmological constant has been considered earlier in the literature \cite{Gibbons:1978ji, Christensen:1979iy,Kazakov:1997nq}. One can extract the precise logarithmic running from these results. The order of magnitude of the anomalous dimensions is expected to be $G\Lambda$. If the UV cutoff is $M_{0}$ then the vacuum energy is of order $M_{0}^{4}$, $\Lambda$ is of the order $M_{0}^{4}/M_{p}^{2}$, and  the Hubble scale is  $H= M_{0}^{2}/M_{p}$. The anomalous gravitational dressings would thus be of order $H^{2}/M_{p}^{2}$. In very early universe, for example if $H$ is of order $0.1M_{p}$ then these estimates suggest that $\g$ and the slow-roll parameter  would be of order $0.01$. One can thus obtain slow-roll inflation  driven entirely by slowly decaying vacuum energy through the nontrivial effective classical dynamics of the Omega field. This provides an example of `inflation without inflaton' or what is termed `omflation' in \cite{Dabholkar:2015qhk,Dabholkar:2015b}. A high value of $H$ is ruled out by current bounds on primordial gravitational waves but it is interesting that an alternative mechanism for inflation is possible. 
It is worth exploring if this mechanism can be embedded in a realistic cosmology and if there are other ways to enhance anomalous dressing, for example, by thermal effects.
A systematic analysis of these effects will be presented in  \cite{Dabholkar:2015b, Bautista:2015d}.

In the present era,  the value of $\g$  is expected to be of order $G\Lambda$, which would be too small to be observationally interesting. Nevertheless, one is led to the conclusion that, in principle, the dark energy will eventually decay even if it happens extremely slowly. 
At a purely theoretical level, this  provides a new perspective on quantum gravity in de Sitter spacetime. 
In a quantum theory of gravity there are no gauge-invariant local observables because the metric itself is dynamical. An important question, independent of any ultraviolet completion of Einstein gravity, is to define gauge invariant observables for the three possible asymptotically maximally symmetric spaces with zero, negative, and positive curvature.
In asymptotically Minkowski spacetime, the observables are the S-matrix elements. In asymptotically Anti de Sitter spacetime the observables are the boundary correlation functions. In both cases, string theory provides  a consistent framework and precise prescriptions for obtaining well-defined finite answers for these quantities. 
On the other hand, quantum gravity in asymptotically de Sitter spacetime presents a number of conceptual difficulties. Considerations of de Sitter entropy \cite{Figari:1975km,Gibbons:1977mu} suggest that the Hilbert space might be finite-dimensional \cite{Banks:1984cw,Bousso:1999cb, Bousso:2000nf} but it is not entirely clear what the gauge-invariant observables on this Hilbert space might be \cite{Hellerman:2001yi,Witten:2001kn,Goheer:2002vf}. Moreover, it has  proved to be difficult to accommodate asymptotically de Sitter spacetime within the framework of string theory. According to a  no-go theorem,  it  is impossible to obtain a de Sitter compactification in the classical supergravity limit \cite{deWit:1986mwo,Maldacena:2000mw}; and all  known constructions in string theory \cite{Kachru:2003aw,Douglas:2006es,Silverstein:2007ac} correspond to metastable vacua which decay nonperturbatively.

Our results imply that perhaps it is not necessary to try to make sense of asymptotically de Sitter spacetime if  we take quantum effects into account even at the perturbative level.    If $\gamma$ is positive, the universe   will be asymptotically Minkowski in the future.  Hence the Penrose diagram looks like  a `house'  with a  sloping roof in the asymptotic future and with a floor on the initial spacelike surface in the asymptotic past where one must impose  an  appropriate cutoff to avoid the initial singularity. In this case,  the observables are the usual cosmological observables computed for a given  wavefunction of the universe defined on the initial value surface.

\appendix

\section{Nearly Static Coordinates}

Static coordinates of de Sitter spacetime are useful for studying the thermodynamic properties of the spacetime \cite{Birrell:1982ix,Spradlin:2001pw}. 
Even thought there is no global timelike Killing vector in de Sitter spacetime, the static coordinates provide a timelike future-oriented Killing vector in the static patch. 
These are also the  natural coordinates for a Schwarzschild-de Sitter solution. Is there an analog of the static coordinates for our new solution?

Since our solution violates the de Sitter symmetry, we do not expect exactly static coordinates. Indeed, it can be shown that no such exact static coordinates exist for our cosmological solution \eqref{modified-deSitter}. However, since any two Robertson-Walker metrics are Weyl equivalent, our solution admits nearly static coordinates in which it is conformal to static de Sitter:
\be\label{nearly-static}
ds^2= \left( \frac{e^{2h T}}{1-h^2R^2}\right)^{\frac{\g}{\g-2}}\, \left[-(1-h^2 R^2)\,dT^2+\frac{1}{1-h^2R^2}\, dR^2 + R^2 \,d\O_2^2 \right]\, ,
\ee
where constant $h$ is defined as
\be
h:=  \frac{2-\g}{2}\, H_* \, a_*^{\g/2}\, .
\ee
In the limit of vanishing $\g$, one recovers  the static de Sitter metric.  

\section{Green Function in Minkowski Spacetime \label{Green}}

The quartic Green equation \eqref{green-equation} is Weyl invariant. Hence the Green function in a conformally flat spacetime is the same as the Green function in Minkowski spacetime:
\begin{equation} 
G( t, \vec{x}; t', \vec{x'}) = \int \frac{dE}{2\pi}\frac{d^3k}{(2\pi)^3} \left(\frac{1}{(E^2-k^2)^2}\right)e^{-iE\Delta t + i\vec{k}.\Delta \vec{x}} 
\end{equation}
%
where $\Delta t := t-t'$ and $\Delta \vec{x} = \vec{x} -\vec{x'}$.
As usual that  the retarded Green function is obtained by lowering  the poles off the real axis in the $E$ integral.  The $E$ integral gives
\begin{equation} \int dE \frac{e^{-iE\Delta t}}{(E-k)^2(E+k)^2} = \frac{2\pi i}{4} \theta(\Delta t) \left[\frac{i\Delta t}{k^2} \left(e^{-ik\Delta t}+e^{ik\Delta t}\right) + \frac{1}{k^3} \left(e^{-ik\Delta t}-e^{ik\Delta t}\right)\right] \end{equation}
The angular integrals can be readily performed to obtain
\begin{equation} \int_0^{2\pi} \frac{d\phi}{2\pi} \int_0^{\pi} \frac{d\theta}{2\pi} \sin(\theta)\ e^{i \vec{k}.\vec{x}} = \frac{1}{2\pi} \int_{-1}^1d(\cos\theta) e^{ik|\Delta \vec{x}|\cos\theta} = \frac{1}{2\pi i |\Delta \vec{x}|}\left(\frac{e^{ik|\Delta \vec{x}|}-e^{-ik|\Delta \vec{x}|}}{k}\right)
\end{equation}
and we are left with the integral
\begin{equation}
\begin{split}
G_{ret}( t, \vec{x}; t', \vec{x'}) =  \frac{\theta(\Delta t)}{8\pi |\Delta \vec{x}|}\int_0^{\infty}\frac{dk}{2\pi} & \left[\frac{1}{k^2}\left(e^{ik|\Delta \vec{x}|}-e^{-ik|\Delta \vec{x}|}\right)\left(e^{-ik\Delta t}-e^{ik\Delta t}\right) + \right. \\
 & + \left. \frac{i\Delta t}{k} \left(e^{ik|\Delta \vec{x}|}-e^{-ik|\Delta \vec{x}|}\right)\left(e^{-ik\Delta t}+e^{ik\Delta t}\right) \right]
\end{split}
\end{equation}
Since the integrand is even, one can extend the integral: $\int_0^{\infty}dk\ \rightarrow\ \half\int_{-\infty}^{\infty}dk$. The integrand has no poles so one can deform the contour slightly below the real axis and write the integrand as  a sum of exponentials with  poles
 \begin{equation}
\begin{split}
G_{ret}( t, \vec{x}; t', \vec{x'}) & =  \frac{\theta(\Delta t)}{16\pi |\Delta \vec{x}|}\int_{-\infty}^{\infty}\frac{dk}{2\pi}  \left[\frac{1}{k^2}\left(e^{ik(|\Delta \vec{x}|-\Delta t)}-e^{ik(|\Delta \vec{x}|+\Delta t)}-e^{ik(-|\Delta \vec{x}|-\Delta t)}+e^{ik(-|\Delta \vec{x}|+\Delta t)}\right)  \right. \\
& + \left. \frac{i\Delta t}{k} \left(e^{ik(|\Delta \vec{x}|-\Delta t)}+e^{ik(|\Delta \vec{x}|+\Delta t)}-e^{ik(-|\Delta \vec{x}|-\Delta t)}-e^{ik(-|\Delta \vec{x}|+\Delta t)}\right) \right]
\end{split}\end{equation}
This can be readily evaluated using the Cauchy residue theorem to obtain
\begin{equation} G_{ret}( t, \vec{x}; t', \vec{x'}) =\frac{\theta( t-t' - | \vec{x} - \vec{x'}|)}{8\pi} \, .\end{equation}
One can verify explicitly that it satisfies the Green equation:
\begin{equation} (\partial_t^4-\nabla_{x}^2\nabla_{x}^2)\frac{\theta(t-r)}{8\pi}= \frac{1}{8\pi}\left[2\delta(t-r) \nabla^2\left(\frac{1}{r}\right)\right] = \delta(t-r)\delta^{(3)}(\vec{x}) = \delta(t)\delta^{(3)}(\vec{x}) \end{equation}
Note that this  retarded Green function receives contribution not only from the points on the light cone but also from the ``wake''  inside the light cone in four dimensions. This phenomenon occurs also in  two dimensions for the Green function of the d'Alembertian which is given by
\be
G^{(2)}_{ret}( t, {x}; t', {x'})= \half \theta( t -t'- | x- x'|)
\ee
It is interesting to note that the Green function for the d'Alembertian receives contributions from within the light-cone for all odd dimensions but gets contributions from points precisely on the light cone in all even dimensions except in two dimensions \cite{Economou:2006gr}. 
\subsection*{Acknowledgments}

A major part of this work was conducted within the framework of  the ILP LABEX (ANR-10-LABX-63)  supported by French state funds managed by the Agence National de la Recherche within the Investissements d'Avenir programme under reference ANR-11-IDEX-0004-02,  and by the project QHNS in the program ANR Blanc SIMI5. 

TB and AG thank the HECAP group at the ICTP for hosting their visits.  AD  thanks the Department of Theoretical Physics at the Tata Institute of Fundamental Research where this work was initiated; and acknowledges the hospitality of the Aspen Center for Physics,  the Benasque Center  and the Theory Division at CERN  during part of this work. 

We thank  Paolo Creminelli, Alba Grassi, Jeff Harvey, and Ashoke Sen for useful discussions. 

\bibliographystyle{JHEP}
\bibliography{weyl}
\end{document}